# Investigation of Nuclear Structures of Self-conjugate Zn, Ge, Se, Kr, Sr Nuclei

Öz-eşlenik Zn, Ge, Se, Kr, Sr Çekirdeklerinin Nükleer Yapılarının İncelenmesi

**Serkan AKKOYUN[1], Tuncay BAYRAM[2]**

[1]*Sivas Cumhuriyet University, Faculty of Science, Department of Physics, Sivas, Turkey*

[2]*Karadeniz Technical University, Faculty of Science, Department of Physics, Trabzon, Turkey*

**Abstract**

Nuclear structures of the atomic nuclei can be theoretically investigated by using nuclear shell model. Generally, a doubly closed-shell nucleus has been considered as inert core and the nucleons outside the core are taken into account in the calculation. It is assumed that the nucleons in the inert core do not move but each valance nucleon out of the core moves under an average potential created by the others. The self-conjugate (N=Z) moderate mass nuclei region is one of the region for the investigation of several phenomena because of the maximum spatial overlap of neutrons and protons. In this study, the nuclear structures of the moderate mass N=Z have been analyzed in the scope of the nuclear shell model by using KSHELL computer code. In the calculations, doubly magic $^{56}$Ni were taken as core and $p_{3/2}$, $f_{5/2}$ and $p_{1/2}$ single particle orbits were used as valance orbits. Different two-body interactions have been taken into account. The results have been compared with each other and the available values existing in the literature.

**Keywords**: Nuclear shell model, nuclear structure, N=Z nuclei.

**Öz**

Atom çekirdeklerinin nükleer yapıları, nükleer kabuk modeli kullanılarak teorik olarak incelenebilir. Genel olarak, bir çift kapalı kabuk çekirdeği kor çekirdek olarak ele alınır ve kor dışındaki nükleonlar hesaplamada dikkate alınır. Kor çekirdekteki nükleonların hareket etmediği, ancak kor dışındaki valans nükleonlarının, diğerleri tarafından yaratılan ortalama bir potansiyel altında hareket ettiği varsayılmaktadır. Öz-eşlenik (N=Z) orta ağırlığa sahip kütleli çekirdeklerin bölgesi, nötronların ve protonların azami uzaysal çakışması nedeniyle, bazı nükleer olguların araştırılması için uygun bölgelerden birisidir. Bu çalışmada, orta ağırlıktaki N=Z çekirdeklerinin nükleer yapı özellikleri, nükleer kabuk modeli kapsamında KSHELL bilgisayar kodu kullanılarak araştırılmıştır. Hesaplamalarda kor çekirdek olarak çift sihirli $^{56}$Ni ele alınmış ve valans orbitalleri olarak $p_{3/2}$, $f_{5/2}$ ve $p_{1/2}$ tek parçacık seviyeleri kullanılmıştır. Farklı iki cisim etkileşmeleri ele alınmıştır. Sonuçlar birbirleriyle ve literatürdeki mevcut değerlerle karşılaştırılmıştır.

Anahtar Kelimeler: Nükleer kabuk modeli, nükleer yapı, N = Z çekirdekler.

## 1. INTRODUCTION

Many nuclear properties of nuclei can be obtained from the mean-field approximation in which proton and neutrons move independently from each other in a common potential. The field is generated as a result of the interaction among nucleons [1-5]. Atomic nuclei at N=Z line at the nuclidic chart are very interesting in nuclear structure studies due to protons and neutrons (nucleons) occupy same orbits. In order to investigate the nuclear structure including N=Z nuclei, the nuclear shell model (SM) is a one of the successful model [6-9]. Several phenomena can be predicted by using this model such as excited energy states, spin/parity of energy levels, electric/magnetic transition probabilities and shapes of the nuclei. In this model, protons and neutrons occupy the single particle orbits from lowest energy state to higher ones leading to the concept of shell structure and shell closures. The nuclei in the moderate mass region can be studied easily via SM [10-12]. Identification of the energy levels of nuclei is an important issue for both experimentalist and theoreticians. By this way, theoretical models can be improved by the comparison of the experimental values while the experimental results can be foreseen. Nuclear SM is similar to the electronic shell model of the atoms. Analogously, valance nucleons in the nuclei which are located out of closed-shells play important roles in the determination of nuclear properties. The nuclei having the magic numbers (2, 8, 20, 28, 50, 82, 126) for proton or/and neutron are named as closed-shell nuclei and these nuclei are used as inert core in the calculations. Closed-shell nuclei are very stable and have completely different properties comparing with their neighbors. It is assumed that the nucleons in the inert core do not move, whereas each valance nucleon out of the core moves under the average potential by the others. Therefore, only valance nucleons are taken into account in the calculations.

Corresponding Author: Serkan AKKOYUN sakkoyun@cumhuriyet.edu.tr

For the SM calculations, there are many computer codes exist in the literature such as KSHELL [13], NuShell [14], Redstick [15], Bigstick [16], Antoine [17] and Oxbash [18]. In this paper, the first $2^+$ and $4^+$ excited state energies, reduced electric transition probabilities (B(E2) from $0^+$ to $2^+$ state, quadrupole deformation parameters ($\beta$), $R_{4/2}$ ratios of moderate mass N=Z nuclei have been calculated and investigated by using KSHELL shell model code [13]. This code enables us to perform nuclear shell-model calculations with M-scheme representation with the thick-restart Lanczos method. The code is easily used on a Linux PC with a many-core CPU and OpenMP library. The code can compute energy levels, spin, isospin, magnetic and quadrupole moments, E2/M1 transition probabilities and one-particle spectroscopic factors. Up to tens of billions M-scheme dimension is capable, if enough memory is available on the computers. In the calculations of the present sturdy, doubly magic $^{56}$Ni nuclei were taken as inert core and $p_{3/2}$, $f_{5/2}$ and $p_{1/2}$ single particle orbits form the valance orbits. Two different Hamiltonians for two-body interactions have separately been used. The results have been compared with each other and the available values in the literature.

## 2. MATERIAL AND METHOD

The model space above the $^{56}$Ni closed shell is suitable for the investigation of moderate mass N=Z nuclei. The model space used in the calculations consists of $p_{3/2}$, $f_{5/2}$ and $p_{1/2}$ valence orbits. The valance nucleons can be located in these three orbits randomly. Because, the nucleons in the core with J=0 do not move from the core and the valance nucleons do not move into the core, we did not consider the nucleons in the closed-shells. In Figure 1, we have illustrates the model space and core for protons and neutrons according to shell model theory. The Hamiltonian of the valance nucleons is given by the following equation,

$$H = E_0 + \Sigma\varepsilon_i + \Sigma<ab;|V|cd;JT> \qquad (1)$$

where $E_0$ is the energy of the inert core, $\varepsilon_i$ is single particle energies of the valance orbits and the last term <ab;JT|V|cd;JT> is two-body interaction among the valance particles. For two-body interaction, *f5pvh* [19] and *jun45* [20] interaction Hamiltonians have been used separately in the calculations. The single particle energies for *f5pvh* interaction are -10.27 MeV, -9.42 MeV and -9.05 MeV for $p_{3/2}$, $f_{5/2}$ and $p_{1/2}$ orbitals, respectively. The interaction Hamiltonian is defined by a set of 63 two-body matrix elements. For *jun45* interaction, the single particle energies are -9.83 MeV, -8.71MeV and -7.84 MeV for $p_{3/2}$, $f_{5/2}$ and $p_{1/2}$ orbitals. Although this interaction also includes $g_{9/2}$ orbit above the $p_{1/2}$, we have not taken into account this orbit for the exact comparison with the former one. The full interaction file contains 334 two-body matrix elements.

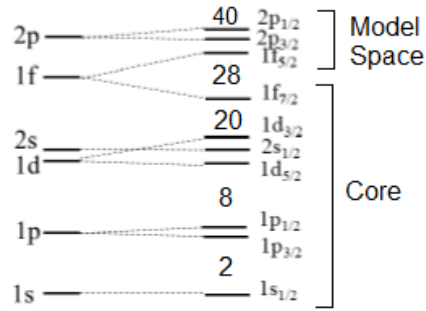

**Figure 1.** Single-particle levels of nucleons grouped to form model space above the $^{56}$Ni core for the calculations

## 3. RESULTS AND DISCUSSIONS

In Figure 1 and 2, we have shown the first $2^+$ and $4^+$ state energies from SM calculations for the investigated $^{60}$Zn, $^{64}$Ge, $^{68}$Se, $^{72}$Kr and $^{76}$Sr nuclei (N=Z) in comparison with the experimental data [21]. In order to analyze nuclear structure, these levels have to be known accurately. The first $2^+$ energies of nuclei can be used as an indicator of collective behavior. Generally, light nuclei have higher energy values for the first $2^+$ state. While the mass numbers of nuclei increase, the $2^+$ energy values decrease. Until atomic number 34, this behavior is seen in this work. From our calculations, the results for $2^+$ energy values stay nearly constant after atomic number 34 for *jun45* interaction and slightly increase for *f5pvh* interaction. For $^{60}$Zn isotope, first $2^+$ excited level energies from different interactions are compatible with the experimental value. But *f5pvh* gives closer value to the experimental value. For $^{64}$Ge isotope, the result from *f5pvh* is still close to the experimental value, but *jun45* gives result far from it. The result of the calculations made using the *f5pvh* interaction and the experimental value are very close to each other for the $^{68}$Se isotope, whereas the calculated result by *jun45* interaction is very far from them. For these three isotopes, theoretical results from both interactions are lower than the experimental values. For $^{72}$Kr isotope, the result is still lower than the experimental value for *jun45* interaction, but *f5pvh* gives larger value. For the last investigated N=Z isotope $^{76}$Sr, the first $2^+$ experimental energy value is very low and not certainly determined in the literature. This corresponds to deformed shell gap at the nucleon number 38. The theoretical value from *jun45* is closer to this low value. Besides, both theoretical results are seen as larger than the low experimental value.

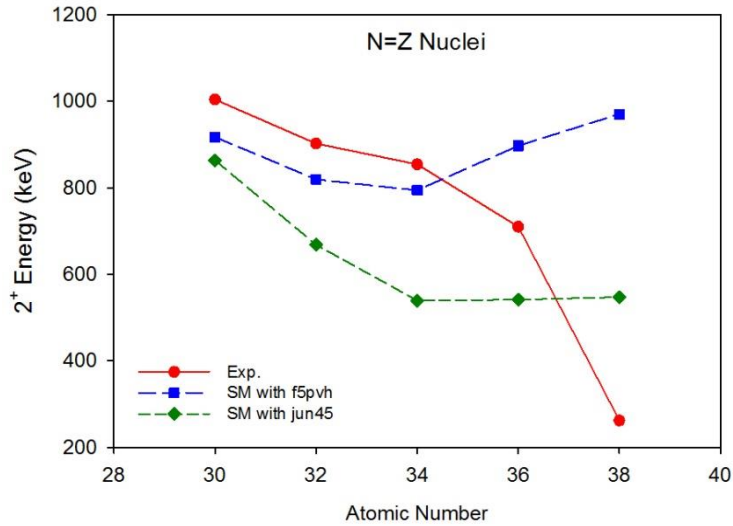

**Figure 2.** First $2^+$ excited state energies for $^{60}$Zn, $^{64}$Ge, $^{68}$Se, $^{72}$Kr and $^{76}$Sr isotopes.

In Figure 2, we have shown first $4^+$ excited state energies for the isotopes under investigation. For $^{60}$Zn isotope, result from *f5pvh* interaction is very close to experimental value. For $^{64}$Ge and $^{68}$Se isotopes, *f5pvh* gives almost same value as the experiment. But after these isotopes, *jun45* starts to give results closer to the experimental data. For $^{72}$Kr isotope, *jun45* value is very close to the experimental value. But for $^{76}$Sr, as in the $2^+$ state, results from the both theoretical calculations are not compatible with the experiment. Again, there is ambiguity in the experimental value at this state for $^{76}$Sr isotope.

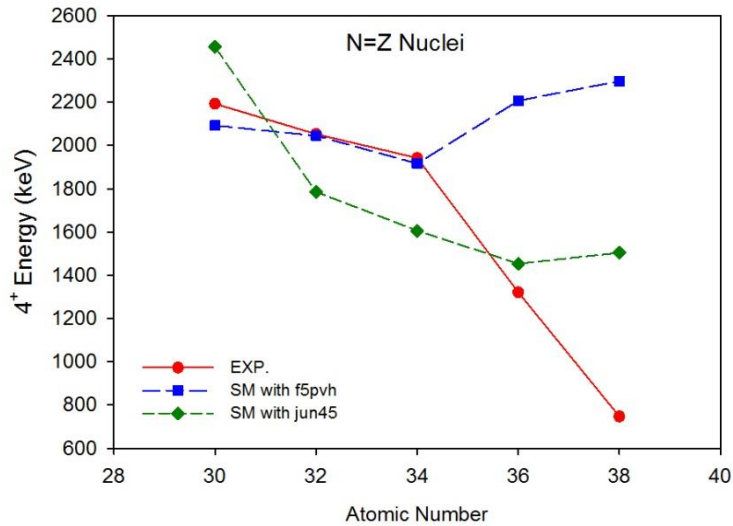

**Figure 3.** First $4^+$ excited state energies for $^{60}$Zn, $^{64}$Ge, $^{68}$Se, $^{72}$Kr and $^{76}$Sr isotopes.

We have also calculated the ratios of $2^+$ to $4^+$ energies ($R_{4/2}$) for the investigated nuclei (Figure 4). This ratio can take different value as < 2.0, ~2.0, ~2.5 and ~3.3 for non-collective, spherical-vibrator, transitional and rigid-rotor structured nuclei, respectively. From $^{60}$Zn to $^{68}$Se, both theoretical with *f5pvh* interaction and experimental results are close to each other and they indicate that these nuclei are transitional character. For $^{72}$Kr and $^{76}$Sr, theoretical results with *f5pvh* interaction show that these nuclei are also transitional, whereas according to the experimental values these are close to rigid-rotor. By using jun45 interaction in the calculations, we have obtained slightly larger values. The ratios are between transitional and rigid-rotor character.

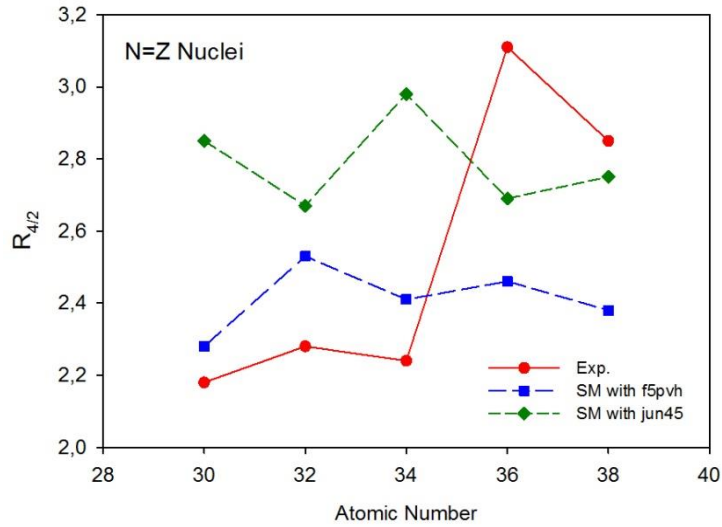

**Figure 4.** $R_{4/2}$ ratios for $^{60}$Zn, $^{64}$Ge, $^{68}$Se, $^{72}$Kr and $^{76}$Sr isotopes.

We have also calculated reduced electric transition probability from the ground state to first $2^+$ state B(E2). Both theoretical results are in good agreement for $^{64}$Ge and $^{68}$Se with the available adopted values in the literature [22]. For $^{60}$Zn, there is no adopted value in the literature. For $^{72}$Kr and $^{76}$Sr, the theoretical results stay nearly constant, but the adopted values start drastically to increase which shows increase of collectivity. We have also given the B(E2) values from other theoretical models in the figure. As is seen in the figure that only HFB+BCS results are in harmony with the both SM calculations.

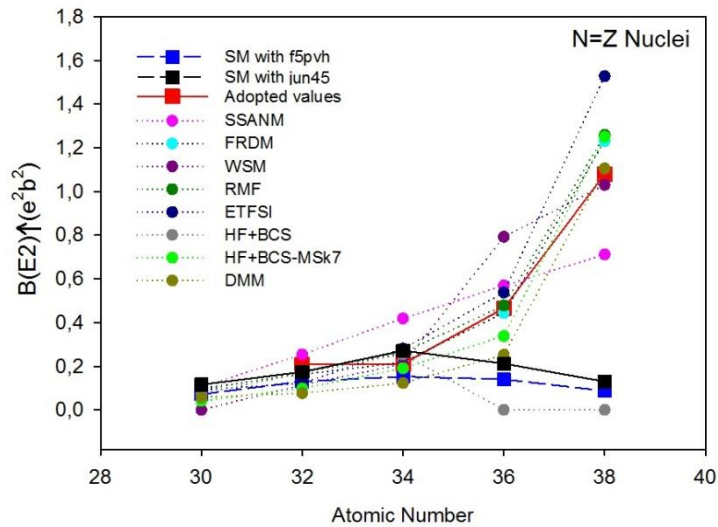

**Figure 5.** B(E2) ratios for $^{60}$Zn, $^{64}$Ge, $^{68}$Se, $^{72}$Kr and $^{76}$Sr isotopes.

Finally in Figure 6, we have shown the quadrupole deformation parameters for the nuclei. Positive value of $\beta_2$ is related with prolate shape while negative value of $\beta_2$ corresponds to oblate shape of nuclei. As can be seen in the figure that all the nuclei under investigation have prolate shape. There is no adopted value for $^{60}$Zn nuclei in the literature. According to the adopted values, $^{72}$Kr and $^{76}$Sr are very deformed nuclei, but for the both theoretical calculations by *f5pvh* and *jun45* interactions, these are less deformed. Additionally, according to the common literature result [23], $^{68}$Se nucleus has oblate shape.

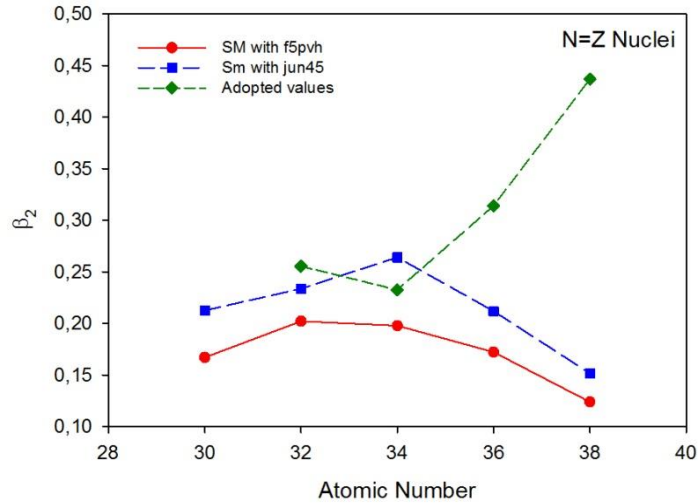

**Figure 6.** Quadrupole deformation parameters ($\beta_2$) for $^{60}$Zn, $^{64}$Ge, $^{68}$Se, $^{72}$Kr and $^{76}$Sr nuclei

## 4. CONCLUSIONS

In this study, nuclear structures of $^{60}$Zn, $^{64}$Ge, $^{68}$Se, $^{72}$Kr and $^{76}$Sr moderate mass N=Z nuclei have been investigated within the nuclear shell model by calculating first $2^+$ and $4^+$ energies, the ratios between these energies, reduced electric transition probabilities from ground state to first $2^+$ state and quadrupole deformation parameters. KShell computer code was used for the calculations on these isotopes with two different two-body interaction Hamiltonians (*f5pvh* and *jun45*). The results have been compared with each other and the available literature and experimental data. For $^{60}$Zn, $^{64}$Ge, $^{68}$Se nuclei, the results from theoretical calculations are in harmony with the literature data. *F5pvh* interaction generally gives closer results than *jun45* interaction for these nuclei. Furthermore, the B(E2) values are theoretically calculated in this study whose data is not exist in the literature.

**Acknowledgement**

This work is supported by the scientific research project fund of Sivas Cumhuriyet University under the project number F-616.